\documentclass[graybox, secnum]{svmult}

\usepackage{mathptmx}       
\usepackage{helvet}         
\usepackage{courier}        
\usepackage{type1cm}        
\usepackage{aas_macros}
\usepackage{subfig}
\usepackage{amsmath}

\usepackage{makeidx}         
\usepackage{graphicx}        
\usepackage{multicol}        
\usepackage[bottom]{footmisc}
\usepackage{hyperref}        
\usepackage{soul}            
\hypersetup{colorlinks=true,urlcolor=blue}

\usepackage[square,numbers]{natbib}
\makeindex             

\renewcommand{\d}{\mbox{d}}
\renewcommand{\P}{\mathcal{P}}
\newcommand{\Ph}{\mathrm{Ph}}
\newcommand{\M}{\mathcal{M}}
\newcommand{\sm}{\textrm{sm}}
\newcommand{\obs}{\textrm{obs}}
\newcommand{\src}{\textrm{src}}
\newcommand{\bkg}{\textrm{bkg}}
\newcommand{\corr}{\textrm{corrected}}
\newcommand{\eff}{\textrm{eff}}
\newcommand{\mdp}{\textrm{MDP}}

\begin{document}
\title*{Analysis of the data from photoelectric gas polarimeters}

\author{Fabio Muleri \thanks{corresponding author}}

\institute{Fabio Muleri \at INAF-IAPS, Via del Fosso del Cavaliere 100, 00133 Rome (Italy) \email{fabio.muleri@inaf.it}}

\maketitle

\abstract{This chapter is dedicated to the description of the tools and procedures for the analysis of the data collected by X-ray photoelectric gas polarimeters, like the ones on-board the Imaging X-ray Polarimetry Explorer (IXPE). Although many of such tools are in principle common with polarimeters working at other energy bands, the peculiar characteristics and performance of these devices require a specific approach. We will start from the analysis of the raw data read-out from this kind of instruments, that is, the image of the track of the photoelectron. We will briefly present how such images are processed with highly-specialized algorithms to extract all the information collected by the instrument. These include energy, time of arrival and, possibly, absorption point of the photon, in addition to the initial direction of emission of the photoelectron. The last is the quantity relevant for polarimetry, and we will present different methods to obtain the polarization degree and angle from it. A simple method, used extensively especially during the development phase of X-ray photoelectric gas polarimeters, is based on the construction and fitting of the azimuthal distribution of the photoelectrons. While such a method provides in principle correct results, we will discuss that there are several reasons to prefer an analysis based on Stokes parameters, especially when one wants to analyze measurements of real, i.e., not laboratory, sources. These are quantities commonly used at all wavelengths because they are additive, and then operations like background subtraction or the application of calibration are trivial to apply, and they are normal and independent variables to a large extend. We will summarize how Stokes parameters can be used to adapt current spectroscopy software based on forward folding fitting to perform  spectro-polarimetry. Moreover, we will derive how to properly associate the statistical uncertainty on a polarimetry measurement and the relation with another statistical indicator, which is in the minimum detectable polarization.}

\section{Keywords}
X-ray; instrumentation; polarimetry; polarimeters; photoelectric effect; data analysis

\section{Introduction: Photoelectric polarimeters}

\begin{figure}
	\centering
	\includegraphics[width=6cm]{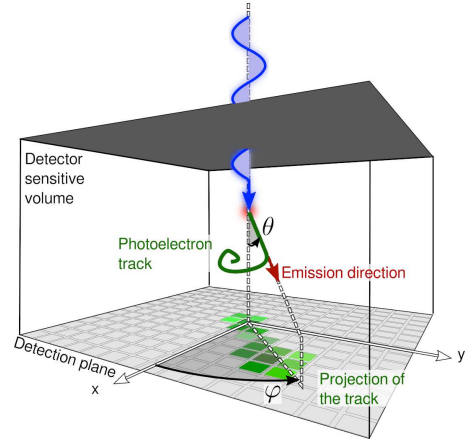}
	\caption{Concept of operations of a typical photoelectric polarimeter. The X-ray photon is absorbed in the sensitive volume, typically filled with a gas mixture. The path of the photoelectron in the medium is the response of the device, which is analyzed off-line to determine the azimuthal direction of emission $\varphi$ which is correlated with the electric field direction of the absorbed photon.}
	\label{fig:PhSketch_v2}
\end{figure}

Photoelectric polarimeters are instruments which measure the polarization in the X-ray energy band using the photoelectric effect. For each detected photon, the detector provides an estimate of the direction of emission of the photoelectron emitted as a consequence of the absorption of the photon. This is the quantity correlated to the polarization of incident radiation and it is conventionally identified by means of its azimuthal and polar angles, $\phi$ and $\theta$, respectively (see Figure~\ref{fig:PhSketch_v2}). The dependence on polarization is expressed by the dependence on $\phi$ and $\theta$ of differential cross section of the interaction, which is \cite{Heitler1954}:
\begin{equation}
	\frac{\d\sigma_{\Ph}}{\d\Omega} \propto
	\cos^2\phi\,\frac{\sin^2\theta}{\left(1-\beta\cos\theta\right)^4}\; ,
	\label{eq:dsdO_Ph}
\end{equation}
where $\beta$ is the photoelectron velocity in units of the speed of light $c$.  Equation~\ref{eq:dsdO_Ph} indicates that the interaction more probably occurs when the photoelectron is emitted parallel to the direction of the photon electric field, i.e., when $\phi$ is 0 or $\pi$, and on the plane orthogonal to the incidence direction, that is, $\theta\approx0$, at least at low energy, $\beta\ll1$. Emission is suppressed in the direction orthogonal to the electric field ($\phi=0$), which, however, holds true in exact terms only when the photon is absorbed by an electron in a spherical shell \cite{Ghosh1983}. This is usually the largely-dominant case in the energy range of real photoelectric polarimeters and, therefore, we will assume in the following that Equation~\ref{eq:dsdO_Ph} can be extended to all events.

The discussion above indicates that the information on the polarization of the absorbed photons is contained in the azimuthal distribution of the photoelectron direction of emission, and photoelectric polarimeters can measure it with different approaches. A simple approach is to relate the probability of events hitting two pixels in finely subdivided semiconductor detectors to the emission direction of the photoelectron (see \cite{Michel2008} and references therein). Typically, modern devices absorb X-rays in a gaseous medium to generate tracks with sufficient length which can be resolved in details on a 2-dimensional image, as in the device in Figure~\ref{fig:PhSketch_v2}. Photoelectron image can be generated with the collection of ions produced by ionization by the photoelectron along its path \cite{Costa2001, Bellazzini2006, Bellazzini2007, Black2007} or by imaging the scintillation light generated by recombination of generated ions \cite{Austin1993}. In all cases, an algorithm is applied to the image to estimate the absorption point of the photon and the initial direction of the photoelectron (see Section~\ref{sec:Reconstruction}). Interestingly enough, practical implementations of photoelectric polarimeters are usually not designed to measure the polar angle of the event. In fact this is not sensitive to polarization, albeit, in principle, it could be useful to discard tracks which are emitted towards the detector plane (or away from it) and then they are more difficult to resolve. 

It is worth noting that the angle $\phi$ in Equation~\ref{eq:dsdO_Ph} and the \emph{measured} azimuthal angle, indicated as $\varphi$ in Figure~\ref{fig:PhSketch_v2} are in principle different: $\phi$ is the azimuthal angle measured from the photoelectron electric field on the plane orthogonal to the incident direction, whereas $\varphi$ is the same angle but measured on the plane of the detector and from one of its reference axes. In the common case in which photons are incident orthogonal (or nearly-so) to the detector, $\phi$ and $\varphi$ simply differs of an offset, which is the polarization angle. In the discussion to follow, we will always implicitly assume that such a condition is satisfied. In the more general case of a large-field-of-view instrument in which photons can impinge at large angles, the response of a photoelectric polarimeter depends not only on polarization of the incident photons but also on their direction, and the analysis requires specific tools \cite{Muleri2014}.

\section{Reconstruction of photoelectron track} \label{sec:Reconstruction}

\begin{figure}
	\centering
	\subfloat[\label{fig:ixpe_track_2.3keV}]{\includegraphics[totalheight=5cm]{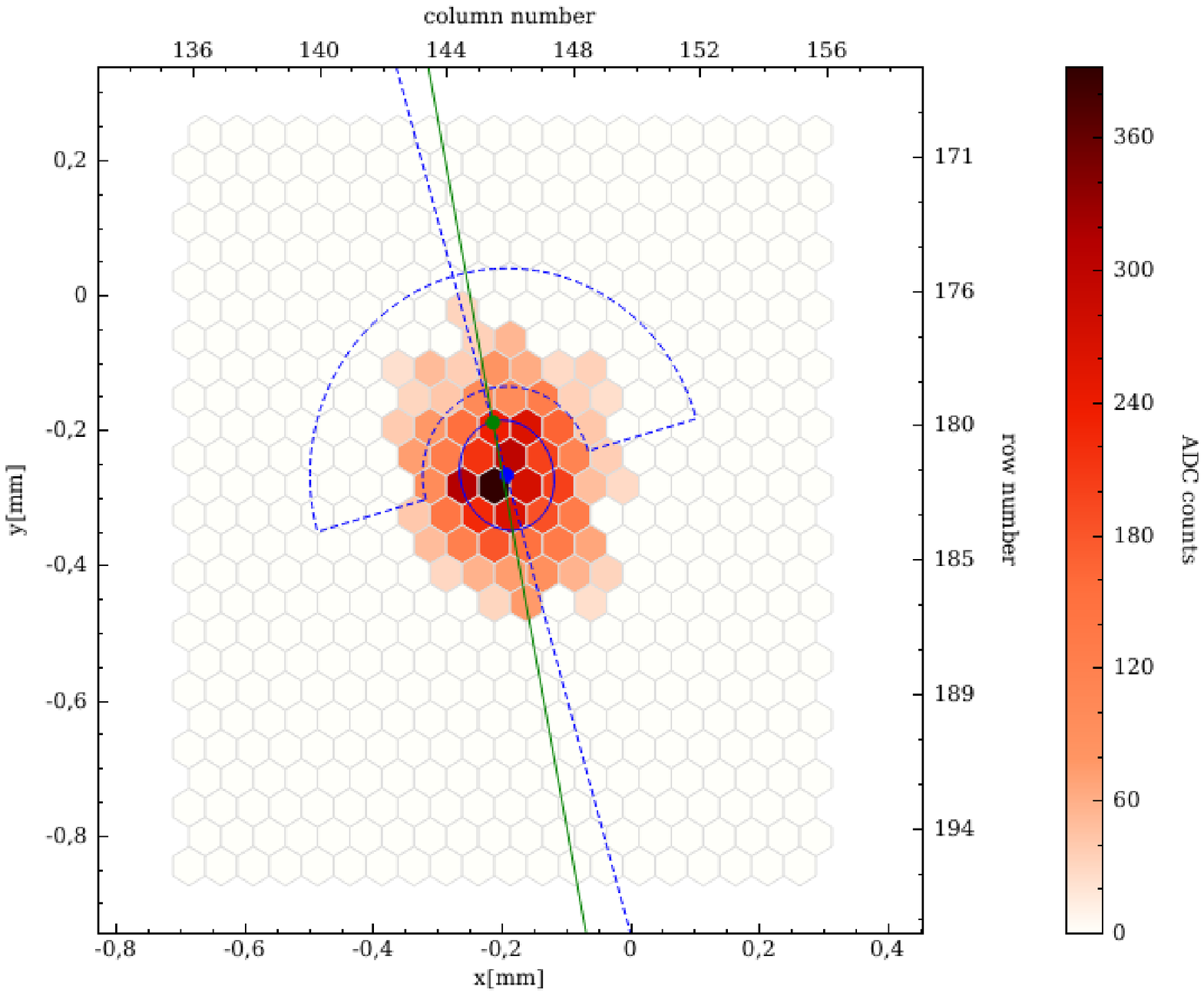}}
	\subfloat[\label{fig:ixpe_track_6.4keV}]{\includegraphics[totalheight=5cm]{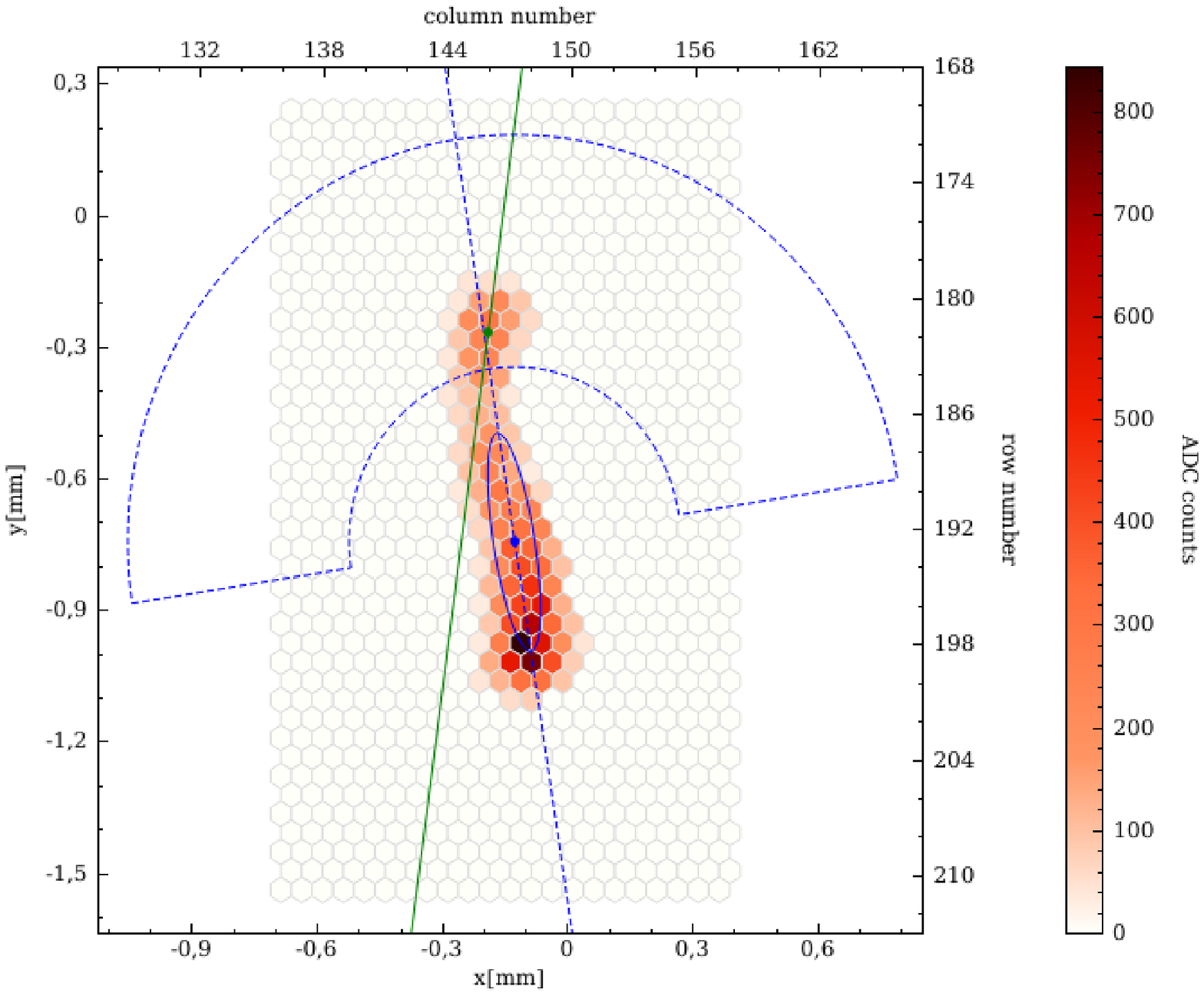}}
	\caption{Photoelectron tracks measured with the photoelectric polarimeters on-board IXPE at 2.3~keV (a) and 6.4~keV (b). The tracks are processed with an algorithm to derive the direction of emission (green line) and the absorption point (green point). Credit: the IXPE team.}
	\label{fig:ixpe_track}
\end{figure}

The response of a photoelectric gas polarimeter is the 2-dimensional image of photoelectron absorbed in its sensitive volume, projected on the plane of the detector (see Figure~\ref{fig:PhSketch_v2}). We show in Figure~\ref{fig:ixpe_track} real photoelectron images generated from X-ray photons absorbed in one of the flight Gas Pixel Detectors\cite{Costa2001, Bellazzini2006, Bellazzini2007} built for the Imaging X-ray Polarimetry Explorer (IXPE, \cite{Weisskopf2016, Soffitta2021, Baldini2021}). Absorbed photons had energy 2.3~keV and 6.4~keV for the events in Figure~\ref{fig:ixpe_track_2.3keV} and Figure~\ref{fig:ixpe_track_6.4keV}, respectively, which lie close to the boundaries of the instrument energy range, which is 2\--8~keV. Photons are absorbed in the region where the track has a lower charge density, as the photoelectron loses more and more energy as it slows down. During its path, the photoelectron suffers scattering with atomic nuclei in the absorbing medium which tend to cancel the information on its initial direction of emission.

As expected, high-energy photons generate longer tracks, which are easier to resolve and shows richer details, whereas the features in low-energy tracks are barely visible. Nonetheless, effective area of X-ray instruments and spectra of astrophysical sources decrease quickly with energy, and, as a matter of fact, peak of sensitivity of photoelectric polarimeters is usually close to the low-energy boundary. The first challenge in the analysis of data from photoelectric polarimeter is therefore to design an algorithm able to estimate the initial direction of the photoelectron, especially when its image shows just a few details.

Different approaches have been put forward to reconstruct photoelectron tracks. The simplest is a custom algorithm originally developed for the GPD \cite{Bellazzini2003}, which is based on the calculation of momenta of photoelectron image after that a clustering algorithm is used to separate the track from noise \cite{Baldini2021}. The barycenter of the charge distribution provides the estimate of the absorption point, and the direction that minimize the second momentum is the direction of the photoelectron. Especially at high energy, when the photoelectron track is longer, it is convenient to repeat the algorithm on the initial part of the track only, which is distinguished as the end with lower charge density. An alternative, which is particularly effective when the track has rich features, is to use more complicate techniques, such as the shortest path problem in graph theory \cite{Li2017} or convolutional neural networks \cite{Kitaguchi2019, Peirson2021}.

The capability of measuring the absorption point of the photon provides to photoelectric polarimeters imaging capabilities. Moreover, the total charge in the photoelectron is proportional to the energy of the absorbed photon. Using the trigger of the event to measure the time of arrival, a photoelectric polarimeter can therefore measure all the information transported by the radiation.

\section{A simple analysis with the modulation curve} \label{sec:ModulationCurve}

\begin{figure}
	\centering
	\includegraphics[width=\textwidth]{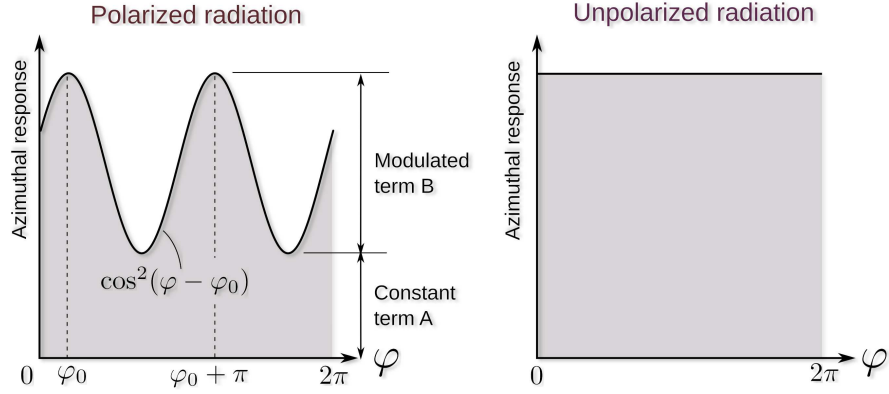}
	\caption{Representation of the response of a photoelectric polarimeter with the modulation curve, that is, the histogram of the photoelectron directions of emission. In case photons are polarized, the modulation curve shows a $\cos^2$ modulation with an amplitude proportional to the degree of polarization and a phase $\varphi_0$ corresponding to the polarization angle. In case of unpolarized radiation, the response is uniform as a function of the azimuthal direction.}
	\label{fig:ModulationCurve}
\end{figure}

The number of photoelectron emitted per azimuthal angle is expected to be anisotropic when absorbed photons are polarized. The signature is specifically a cosine square modulation, remindful of $\phi$-dependence of differential cross section (see Equation~\ref{eq:dsdO_Ph}), with amplitude proportional to the polarization degree and with a maximum in the direction of the polarization angle. To evaluate it, a simple procedure is to build the histogram of the reconstructed photoelectron emission angles with $M$ bins, named \emph{modulation curve} (see Figure~\ref{fig:ModulationCurve}), and fit it with the function\footnote{The modulation curve can be fitted with alternative functions, e.g. $\M(\varphi) = C + M \cos\left[2(\varphi-\varphi_0)\right]$. It is trivial to use trigonometric functions to prove that these treatments are equivalent.}:

\begin{equation}
	\M(\varphi) = A + B \cos^2(\varphi-\varphi_0).
\end{equation}

In absence of background, the polarization degree $\P$ is defined as the ratio between the flux of the polarized component with respect to the total. The first term is causing the observed modulation in the modulation curve, and therefore $\P$ is proportional to the measured \emph{modulation amplitude} $a$: 
\begin{equation}
	a=\frac{\int_{-\pi}^{+\pi}B\cos^2(\varphi-\varphi_0) d\varphi}{\int_{-\pi}^{+\pi}\left[A+B\cos^2(\varphi-\varphi_0)\right] d\varphi}=\frac{B}{B+2A}.
	\label{eq:ModulationCurveClassic}
\end{equation}

The normalization factor between $\P$ and $a$ is named \emph{modulation factor}, which is the modulation measured when absorbed photons are 100\% polarized:
\begin{equation}
	\P = \frac{a}{\mu} = \frac{1}{\mu}\frac{B}{B+2A}.
\end{equation}
The modulation factor of a polarimeter accounts for the fact that real detectors can not provide a perfect reconstruction of the event, that is, the direction of emission of the photoelectron is reconstructed with a certain uncertainty. As the quality of the photoelectron track changes with energy, also the modulation factor does. It can be obtained from detailed simulations or accurate measurements, the latter being necessary for a fully representative result. The polarization angle is the value $\varphi_0$ obtained by the fit with Equation~\ref{eq:ModulationCurveClassic}, which is defined modulo its period $\pi$. 

\section{The Minimum Detectable Polarization}

\begin{figure}
	\centering
	\includegraphics[width=\textwidth]{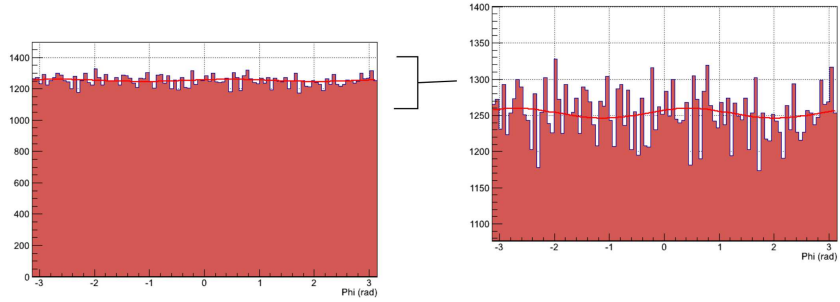}
	\caption{Example of a real modulation curve obtained with unpolarized radiation with the GPD. Zooming in, a certain cosine-square modulation is always measured.}
	\label{fig:SpurMod}
\end{figure}

When the procedure described in the previous section is applied to a modulation curve measured in case of unpolarized photon, the result is always the detection of a certain modulation amplitude, albeit small (see Figure~\ref{fig:SpurMod}). This is caused by the fact that the number of detected photons in a certain bin of the modulation curve is Poisson-distributed, and the independent statistical fluctuations among different bins always cause to detect a cosine-square modulation. The amplitude of such a component is to all extent the statistical limit of the measurement. This is quantified introducing the \emph{Minimum Detectable Polarization} (MDP), which is defined as the maximum polarization which can be produced by statistical fluctuations only in absence of true source polarization, at a certain confidence level $C=99\%$. The MDP depends on the number of source (and background) events, that is, on the instrument effective area, and on the modulation factor. Therefore, the MDP is also a meter of the detector sensitivity, and, once referred to an observation of reference source for a specific duration, it can be used to compare the sensitivity of different experiments.

The value of the MDP for a certain observation for which $N$ events are collected can be derived by the probability $P$ of measuring a certain polarization $\P$ in case the true polarization $\P_0$ is zero, integrated over the observed angle of polarization \cite{Weisskopf2010}:
\begin{equation}
	P(N, \P | \P_0 = 0) = \frac{N}{2} \mu^2 \P \exp\left[-\frac{N}{4}(\mu\P)^2\right].
\end{equation}

\begin{figure}
	\centering
	\includegraphics[width=0.7\textwidth]{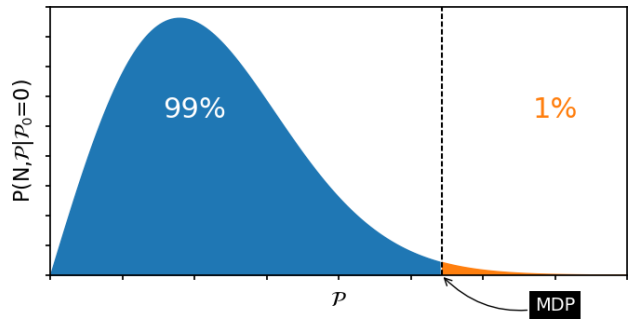}
	\caption{Probability distribution function of the measured polarization degree when the source is unpolarized. The MDP is defined as the value for which the cumulative distribution function is 0.99.}
	\label{fig:mdp}
\end{figure}

By definition, the MDP is the value at which the cumulative distribution function of $\P$ is 99\%, that is (see Figure~\ref{fig:mdp}):
\begin{equation}
	0.99 = \int_0^{_\mdp} P(N, \P | \P_0 = 0)\d\P = 1-\exp\left(-(\mdp^2\;N\;\mu^2)/4\right).
	\label{eq:mdp_integration}
\end{equation}

Inverting Equation~\ref{eq:mdp_integration} for the MDP, one finds:
\begin{equation}
	MDP = \frac{\sqrt{-4\;\ln 0.01}}{\mu\sqrt{N}}\approx\frac{4.29}{\mu\;\sqrt{N}},
\end{equation}
which is the well-know expression in case the background is negligible. Incidentally, this result quantifies the claim that polarimetry is a "photon-hungry" technique. To achieve, with an instrument with average modulation factor 30\%, an MDP of 1\%, that is, smaller than the polarization expected from many astrophyiscal sources, one needs to collect $\sim2\cdot10^6$ events. This is orders of magnitude larger than the statistics needed to build a spectrum.

In case of non-negligible background, the MDP can be calculated with \cite{Elsner2012}:
\begin{equation}
	\mdp \approx \frac{4.29}{\mu}\sqrt{\frac{N_\src+N_\bkg}{N_\src^2}}.
	\label{eq:mdp_bkg}
\end{equation}
It is worth stressing that Equation~\ref{eq:mdp_bkg} is valid when the background contribution is known, intending its expected modulation amplitude and phase. In this case, even if the background can be removed from the measurement, its statistical fluctuations still limits the achievable MDP.

\section{Stokes parameters} \label{sec:StokesParameters}

The approach presented in Section~\ref{sec:ModulationCurve} has the advantage to be simple and close to the underlying physics of the detector, and to provide directly the polarization degree and angle, which are the physical observables which one is usually interested in. However, it shows limitations in practice, for example it is not trivial to subtract a background or apply the calibration of the detector with proper statistical treatment. Manipulation of the measured signal can be easily achieved by using Stokes parameters.

Stokes parameters are a 4-dimension vector, usually indicated with $(I,Q,U,V)$, which is used to completely describe the state of polarization of radiation, intending both its linear and circular polarization and its intensity \cite{Tinbergen1996}. Their use has been common especially at wavelengths longer than X-rays, but as we will see Stokes parameters are adequate and convenient also to describe polarization at higher energy.

There are several equivalent ways to define Stokes parameters, each fitting better with the measurement procedure in a specific energy range. For example, in optical light or infrared, the intensity of radiation component polarized in a certain direction, e.g. at $I_0$ at 0$^\circ$, $I_{90}$ at 90$^\circ$, can be easily measured as it is for right and left circularly polarized components, $I_{rc}$ and $I_{lc}$. Then, Stokes parameters are defined as \cite{Tinbergen1996}:

\begin{equation*}
	\begin{cases}
		I = I_0 + I_{90} \\
		Q = I_0 - I_{90} \\
		U = I_{45} - I_{-45} \\
		V = I_{\textrm{rc}} - I_{\textrm{lc}} \\
	\end{cases} .
\end{equation*}

In radio, one measures the amplitude $E_x$ and $E_y$ of the wave in orthogonal directions as a function of time, and operations like correlation (and conjugation) of the signal are readily available. In this case, Stokes parameters are usually defined as \cite{Tinbergen1996}:
\begin{equation*}
 \begin{cases}
	I = \overline{E_x E_x^* + E_y E_y^*} \\
	Q = \overline{E_x E_x^* - E_y E_y^*} \\
	U = \overline{E_x E_y^* + E_y E_x^*} \\
	V = i\;\overline{E_x E_y^* - E_y E_x^*} \\
 \end{cases} ,
\end{equation*}
where the line indicates the averaged value over time.

In the X-ray energy range, current implementations of polarimeters are not sensitive to circular polarization and in fact only $I$, $Q$ and $U$ can be measured and will be discussed hereafter. A method to calculate them is fitting the modulation curve with a function depending explicitly on Stokes parameters\cite{Strohmayer2013}\footnote{In our definition we inverted $Q$ and $U$ with respect to the the original paper \cite{Strohmayer2013} for consistency with the usual formulae to derive polarization at other wavelengths.}:
\begin{equation}
	\M(\varphi) = I + Q \cos 2 \varphi + U \sin 2 \varphi.
	\label{eq:ModulationCurveStokes}
\end{equation}

An alternative approach consists in calculating the Stokes parameters for each event $(i_k, q_k, u_k)$ \cite{Kislat2015}. If $\varphi_k$ is the (reconstructed) emission angle of the $k$-th photoelectron \footnote{With respect to \cite{Kislat2015}, we added a multiplicative factor equal to 2 to the definition of $q_k$ and $u_k$ (compare our Equation~\ref{eq:stokes_event} to Equation~9 in \cite{Kislat2015}). Although this makes Stokes parameters Q and U variable in the $[-2\div2]$ range rather than in the usual $[-1\div1]$, such a choice allows one to have equation for the calculation of polarization degree consistent with other wavelengths. Notwithstanding, the two treatments are fully equivalent.}:
\begin{equation}
	\begin{cases}
		i_k = 1 \\
		q_k = 2\cos2\varphi_k \\
		u_k = 2\sin2\varphi_k 
	\end{cases} .
	\label{eq:stokes_event}
\end{equation}
From this definition, it follows that Stokes parameters $q_k$ and $u_k$ are the abscissa and the ordinate of a vector with length 2 and forming an angle $2\varphi_k$ with the $x$-axis. Although it is more common at other wavelengths to define them as vector with unit length, the factor 2 takes into account the peculiar $\cos^2$ distribution of events observed with X-ray instrumentation and we will adopt it in the following. 

Stokes parameters for a measurement are obtained by summing event Stokes parameters over the entire data set:
\begin{equation}
	\begin{cases}
		I = \sum\limits_{k=1}^{N} i_k = N \\
		Q = \sum\limits_{k=1}^{N} q_k \\
		U = \sum\limits_{k=1}^{N} u_k 
	\end{cases} .
	\label{eq:stokes_all}
\end{equation}

Stokes parameters can at any moment be converted to (linear) polarization degree and angle (and vice versa). However, the usual formulae applicable at all wavelengths must be modified to account for the modulation factor of the instrument, which is a peculiarity of X-ray polarimeters: 
\begin{eqnarray}
\P &=& \frac{1}{\mu} \frac{\sqrt{Q^2 + U^2}}{I} = \frac{1}{\mu} \sqrt{q^2+u^2} 
\label{eq:p}\\
\varphi_0 &=&\frac{1}{2} \arctan\frac{U}{Q} = \frac{1}{2} \arctan\frac{u}{q}
\label{eq:phi}
\end{eqnarray}
where $q=Q/I$ and $u=U/I$ are \emph{normalized} Stokes parameters.

It is worth noting that event-by-event calculation of Stokes parameter does not provides expected values \emph{identical} to those obtained by fitting with the function $\M(\varphi)$ in Equation~\ref{eq:ModulationCurveStokes}; in fact, the obtained values are proportional, the constant of proportionality being the number of bins in the modulation curve, $M$. Nevertheless, the two approaches are completely equivalent as the observable quantities are the polarization degree and angle, which are invariant for proportional Stokes parameters. In the following, we will assume to determine Stokes parameters with the event-by-event approach, and to use normalized Stokes parameters when convenient to make the measured value independent on the number of collected events. 

We also stress that the normalization by the modulation factor is typically done when calculating the polarization degree with Equation~\ref{eq:p}, but one can equivalently define $\mu$-normalized Stokes parameters:

\begin{equation}
	\begin{cases}
		I = \sum\limits_{k=1}^{N} i_k = N \\
		Q = \frac{1}{\mu}\sum\limits_{k=1}^{N} q_k \\
		U = \frac{1}{\mu}\sum\limits_{k=1}^{N} u_k
	\end{cases} 
	\;\Rightarrow \;
	\begin{cases}
		\P = \frac{\sqrt{Q^2 + U^2}}{I}  \\
		\varphi_0 = \frac{1}{2} \arctan\frac{U}{Q}
	\end{cases} .
\end{equation}

We will not develop further this possibility in the following, sticking to the more conventional definition given in Equation~\ref{eq:stokes_all}. However, the interested reader can easily derive the formulae discussed below in the former assumption.

\section{Properties of Stokes parameters}

The use of Stokes parameters with respect that the approach described in Section~\ref{sec:ModulationCurve} has several advantages. While $\P$ (or $a$) and $\phi_0$ are dependent, Stokes parameters can be considered normal and independent variables if a large number of events are collected and the measured modulation is small \cite{Strohmayer2013, Kislat2015}. Both of these conditions are satisfied in practice for all observations of astrophysical sources carried out with photoelectric polarimeters and their validity will be adopted hereafter. In such an assumption\footnote{The interested reader is referred to \cite{Kislat2015} and \cite{Montgomery2015} for the more general case.}, the standard deviation on the (normalized or not) Stokes parameters is simply:
\begin{eqnarray}
	\begin{cases}
		\sigma_I \approx \sqrt{N} \\
		\sigma_Q \approx \sqrt{2N} \\
		\sigma_U \approx \sqrt{2N}
	\end{cases}\; \textrm{or} \; 
	\begin{cases}
		\sigma_q \approx \sqrt{2/N} \\
		\sigma_u \approx \sqrt{2/N}
	\end{cases} .
	\label{eq:stokes_uncertainty}
\end{eqnarray}

Stokes parameters are additive. In case the observation is affected by a significant background, it is trivial to subtract it once that its Stokes parameters $(I_{\bkg}, Q_{\bkg}, U_{\bkg})$ are known. The first step is to re-scale the background Stokes parameters, assumed to be measured with an observation with duration $T_{bkg}$, to the duration of the observation $T_{\obs}$. Then, if $(I, Q, U)$ is measured for the observation and $\eta=\frac{T_{\obs}}{T_{\bkg}}$, source Stokes parameters are derived with:
\begin{equation}
	\begin{cases}
		I = I_{\src} + \eta\; I_{\bkg} \\
		Q = Q_{\src} + \eta\; Q_{\bkg} \\
		U = U_{\src} + \eta\; U_{\bkg}
	\end{cases} 
	\;\Rightarrow \;
	\begin{cases}
		I_{\src} = I - \eta\; I_{\bkg} \\
		Q_{\src} = Q - \eta\; Q_{\bkg} \\
		U_{\src} = U - \eta\; U_{\bkg}
	\end{cases} .
	\label{eq:stokes_bkg}
\end{equation}
Statistical uncertainty on the source Stokes parameters can be derived by standard error propagation of uncertainty on the measured and background values, e.g.:
\begin{equation}
	\sigma_{Q\_\src} = \sqrt{\sigma_Q^2 + \left(\frac{T_{\obs}}{T_{\bkg}}\right)^2\sigma_{Q\_\bkg}^2} .
\end{equation}

Additivity of Stokes parameters is handy also to apply calibration. While the response of an ideal instrument is, apart from statistical fluctuations of the signal, azimuthally uniform, real devices may present a signal also in case of unpolarized radiation. The second armonic of such a signal, which has a period of 180$^\circ$, is usually referred to as \emph{spurious modulation} and it has the same signature as the signal generated by a genuine polarization. Therefore, spurious modulation has to be subtracted from data to obtain the true source polarization. 

The amplitude and phase of spurious modulation can be measured with accurate calibration measurements, or, if possible, derived from simulations. As the modulation genuinely produced by polarization, spurious modulation can be expressed with Stokes parameters, $(I_{\sm}, Q_{\sm}, U_{\sm})$, and as such subtracted from measured values as a background. One has only to assume $\eta = 1$ in Equations~\ref{eq:stokes_bkg}.

A notable case is when spurious modulation is dependent on photon energy and absorption point. In this case, a convenient approach is to subtract event-by-event spurious modulation so that the subsequent analysis can proceed essentially as for an ideal detector \cite{Rankin2022}. The first step is to use calibrations (or simulations) to generate maps of \emph{normalized} Stokes parameters of spurious modulation at different energies, $(q_{\sm}, u_{\sm})$. The maps are then interpolated to find the estimate of the spurious modulation for the measured absorption point $(\bar{x}, \bar{y})$ and photon energy, $\bar{E}$. The spurious-modulation corrected Stokes parameters of the event are derived by the measured photoelectron angle $\varphi_k$ with:
\begin{equation}
	\begin{cases}
		q_k^{\corr} = q_k - q_{\sm}(\bar{x}, \bar{y}, \bar{E}) = 2\cos 2\varphi_k - q_{\sm}(\bar{x}, \bar{y}, \bar{E}) \\
		u_k^{\corr} = u_k - u_{\sm}(\bar{x}, \bar{y}, \bar{E}) = 2\sin 2\varphi_k - u_{\sm}(\bar{x}, \bar{y}, \bar{E})
	\end{cases} .
\end{equation}
Corrected Stokes parameters are not \emph{proper} Stokes parameters, in the sense that they can not be considered the abscissa and the ordinate of a vector with length 2 as it is for $q_k$ and $u_k$. Practical effect is that, from them, it is not possible to properly defined a \emph{corrected} photoelectron angle $\varphi_k^{\corr}$. Notwithstanding, it can readily be shown  that, when summed over a data set, the expected value of the corrected Stokes parameters is the source values\cite{Rankin2022}:
\begin{equation}
	\begin{cases}
		<Q^\corr> = \sum_k q_k^{\corr} = Q_{\src} = Q - Q_\sm \\
		<U^\corr> = \sum_k u_k^{\corr} = U_{\src} = U - U_\sm
	\end{cases} .
\end{equation}
The standard deviations is approximately the same as the one of the observation, e.g., $\sigma_{Q\_\corr} \approx\sigma_Q$, as long as the calibration of spurious modulation is carried out with an adequate statistics larger than that of the observation. For the more general case, the reader is referred to \cite{Rankin2022}.

Event-by-event Stokes parameters allow to introduce weights in the analysis. These are intended to increase the importance of those tracks which can be reconstructed with a smaller uncertainty because of the better quality of the photoelectron image. Depending on the algorithm used, different parameters can be used to quantify the quality of the track. It can be shown that the best weight is proportional to the modulation factor, and topological properties of the charge distribution or the output of convolutional neural networks have been used \cite{DiMarco2022,Peirson2021,Marshall2021}.

Weights are introduced in the event-by-event analysis with a simple modification of Equation~\ref{eq:stokes_event} \cite{Kislat2015}:
\begin{equation}
	\begin{cases}
		i_k = w_k \\
		q_k = w_k \cdot 2\cos2\varphi_k \\
		u_k = w_k \cdot 2\sin2\varphi_k 
	\end{cases} .
	\label{eq:stokes_event}
\end{equation}

The effect of weights is to increase the modulation factor, at the expense of a reduction of the "effective" number of events used in the analysis. The latter effect can be quantified with the introduction of the quantity:
\begin{equation}
	N_{\eff} = \frac{\left( \sum_k w_k \right)^2}{\sum_k w_k^2} = \frac{I^2}{W_2} \; \textrm{with}\; W_2 = \sum_k w_k^2 .
\end{equation}
In fact, $N_\eff$ replaces the number of collected events $N$ in the formulae for the calculus of sensitivity. In weighted analysis, the standard deviations on normalized Stokes parameters becomes \cite{Kislat2015}:
\begin{equation}
	\sigma_q = \sigma_u \approx \sqrt{\frac{2}{N_\eff}}
\end{equation}
and the MDP (in case of absence of background) is:
\begin{equation}
	\mdp=\frac{4.29}{\mu\sqrt{N_\eff}}.
\end{equation}

Weights allow to increase the overall sensitivity of a photoelectric polarimeter. Typical values is an increase of a few tens of \% in the modulation factor, a reduction of $N_\eff/N\sim0.9$ and an overall decrease of the MDP $>$10\% \cite{DiMarco2022}.

\section{Spectro-polarimetry with Stokes parameters and forward-folding}

Definition and additive property make Stokes parameters essentially flux quantites. $I$ is the source flux, while $Q$ and $U$ encode, in addition to source flux, the polarization angle and degree of the source. As these three quantities completely describe the source emission and they are dependent, it is highly desirable that their analysis is carried out simultaneously. \cite{Strohmayer2017} defined a simple method to extend the common tools developed for X-ray spectroscopy, applicable to $I$, to include also linear polarization represented by $Q$ and $U$. This provides a standardized way of performing spettro-polarimetry and we will review the results in the following.

Spectral analysis is often carried out with software, like XSPEC \cite{Arnaud1996}, which adopts a procedure named \emph{forward folding}. This consists in modeling the source spectrum from the measured one with the known response of the instrument. The first step is to define an input model with certain start parameters. Instrument response functions are used to calculate what the measured spectrum would be in case the source had the assumed model. The process is iterated to find the input model parameters which minimize the difference between the calculated and measured spectra with a procedure such as $\chi^2$ fitting. 

The response function used for spectral analysis comprises the effective area $\epsilon(E)$ of the instrument, which is its collecting area, and the redistribution matrix $R(E_0,E)$. The latter expresses the probability that a photon with energy $E$ is reconstructed with an energy $E_0$. To handle polarization, an additional quantity has to be introduced which accounts for the probability that a photon with a certain polarization is measured with a different one. It can be shown \cite{Strohmayer2017} that such a probability is related to the amplitude of the instrumental response to polarization and, ultimately, it is the modulation factor $\mu(E)$ of the instrument. The underlying assumption is that the polarimetric response is independent on the effective area and energy dispersion, which is typically verified for real instruments, and that it does not depend on the polarization degree or angle of the source. The last assumption holds true when the response of the instrument can be treated as ideal, that is, spurious modulation, if present, is subtracted with a method such as that described above.

The effective area, the redistribution matrix and the modulation factor are all the response functions which are needed to perform the forward folding procedure with spectro-polarimetric data. In fact, in case the model Stokes parameters are $\mathcal{I}, \mathcal{Q}$ and $\mathcal{U}$, the observed Stokes parameters $I, Q$ and $U$ over a certain energy bin can be calculated with:

\begin{equation}
\begin{cases}
I = \int_{E'} \mathcal{I}(E)\;\epsilon(E')\; R(E',E) \d E'  \\
Q = \int_{E'} \mathcal{Q}(E)\;\mu(E')\; \epsilon(E')\; R(E',E) \d E'  \\
U = \int_{E'} \mathcal{U}(E)\;\mu(E')\; \epsilon(E')\; R(E',E) \d E'  \\
\end{cases} .
\label{eq:forward_folding}
\end{equation}

It is worth stressing that in this approach the analysis of $Q$ and $U$ requires ``only'' to use a response function $\mu(E') \times \epsilon(E')$ instead of the canonical $\epsilon(E')$ commonly used for $I$ by current software for spectral fitting. This is a relatively small change in the analysis flow and, needless to say, a great advantage when one wants to use the power of spectral analysis together with polarization.

\section{Polarization and its statistical uncertainty}

\begin{figure}
	\centering
	\includegraphics[width=0.7\textwidth]{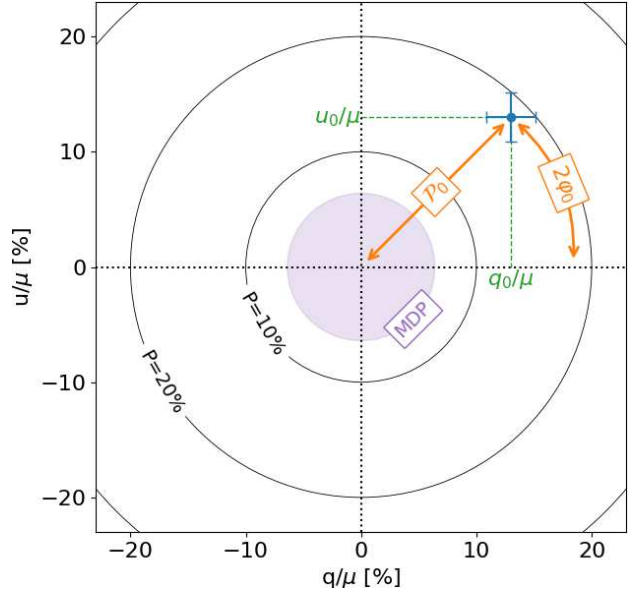}
	\caption{Example of visualization of a measurement result with Stokes parameters. The measured values are $q_0/\mu\approx13$\% and $u_0/\mu\approx13$\%. Modulation factor is assumed to be $\mu=0.3$ and the number of collected counts is 50,000. The 1-$\sigma$ uncertainty on either $q$ or $u$ is $\sigma_q=\sigma_u=\frac{1}{\mu}\sqrt{\frac{2}{N}}$, which is Equation~\ref{eq:stokes_uncertainty} 
but normalized by the modulation factor.}
	\label{fig:stokes_measurement}
\end{figure}

Visualization of the result of a polarization measurement with proper uncertainties can be done in different ways. A first option is to use Stokes parameters, which in this case it is convenient to normalize for both $I$ and the modulation factor $\mu$, to have a result directly expressed in polarization and not in amplitude. An example is reported in Figure~\ref{fig:stokes_measurement}. Measured values are $q_0=4$\% and $u_0=4$\% which, once normalized by $\mu$ (assumed to be 0.3), translate in the coordinates $(\frac{4}{0.3},\frac{4}{0.3})$. The 1-$\sigma$ uncertainty on either $q$ or $u$ is  $\sigma_q=\sigma_u=\frac{1}{\mu}\sqrt{\frac{2}{N}}$, which is Equation~\ref{eq:stokes_uncertainty} but normalized by the modulation factor. The measured polarization degree $\P_0$ is the radius of the measured point from the origin, while the polarization angle is half of the angle that the radius forms with the abscissa. Circles with different radii and centered in the axis origin represent constant polarization values and also the MDP can be represented as a circle.

It is worth stressing here once more the meaning (and difference) between the MDP and the error on the measurement. The former quantity is a very simple statistical quantity, which expresses the maximum polarization that the user should expect to measure (at a confidence level of 99\%) because of statistical fluctuations in the assumption that the source was \emph{not} polarized. If the measured value is larger than the MDP, this means that there is a probability smaller than 1\% that such a value would be measured in case the source was unpolarized, and then the user can claim that the source is polarized at this confidence level. On the contrary,  if the measured value is below the MDP, there is a high chance that the user obtained that value only because of statistical fluctuations and not because the source was genuinely polarized. 

Error on the measurement has a completely different meaning. It expresses how large is the interval in which the true value lies, and then it can be used to compare the measured value with respect to, e.g., a model prediction. So, to be credible, the measured value must be higher than the MDP \emph{and} the error on the measurement must not include the space region in which the polarization is zero, that is, the origin of axis in Figure~\ref{fig:stokes_measurement}. Clearly, these two requirements are interdependent, as both depend essentially on the number of collected counts and the modulation factor, and they should provide a coherent indication on the credibility of the detection. 

There are cases in which it is more convenient to plot polarization degree and angle instead of Stokes parameters. In this case, one has to remember that these two parameters are dependent to properly derive the statistical uncertainty on the measurement. To derive the confidence region of a measurement in the $(\P,\varphi)$ plane,  we start from the assumption that Stokes parameters are normally distributed and independent. As we discussed in Section~\ref{sec:StokesParameters}, this holds true when the $\mu\P\ll1$. In this assumption, the probability distribution function when the measured value is $(q_0,u_0)$ is a bivariate normal distribution \cite{Kislat2015}:
\begin{equation}
	P(N, q | q_0, u | u_0) = \frac{1}{2\pi\sigma^2}\exp\left[ -\frac{(q-q_0)^2 + (u-u_0)^2}{2\sigma^2} \right] ,
	\label{eq:p_stokes}
\end{equation}
where, for uniformity of our notation, Stokes parameters are normalized by $I$ but not by the modulation factor, and then we will have to normalize by $\mu$ when passing from Stokes parameters to polarization. 

\begin{figure}
	\centering
	\includegraphics[width=0.7\textwidth]{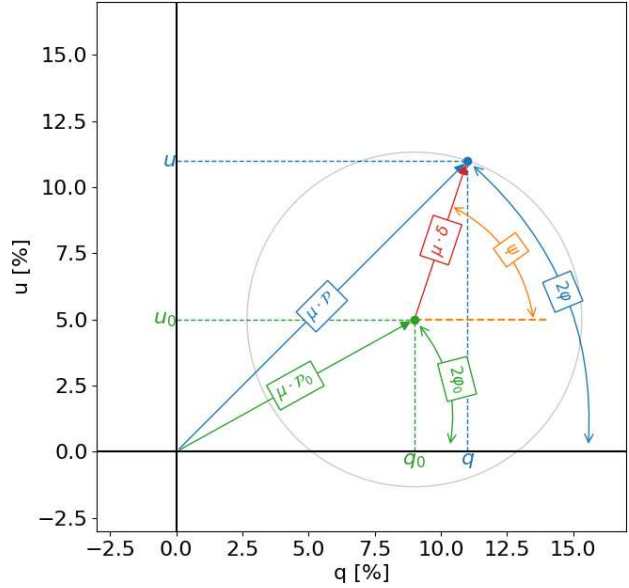}
	\caption{Definition of the quantities introduced for deriving the statistical uncertainty on polarization degree and angle. Measured normalized Stokes parameters are $(q_0, u_0)$, and measured polarization degree and angle $(\P_0, \varphi_0)$. The generic point on the plane is $(q,u)$, or $(\P, \varphi)$. The difference in polarization degree between the generic point and the measured point is $\mu\delta_\P$, }
	\label{fig:exp_term}
\end{figure}

By definition, the statistical uncertainty at a certain confidence level $C$ is the region such that, integrating over it, we obtain $C$:
\begin{equation}
C = \int\int P(N, q | q_0, u | u_0) \d q \d u = \int\int P(N, \P | \P_0, \varphi_0) \d\P\d\varphi.
\label{eq:confidence}
\end{equation}
Now, we have to find a convenient way to express $P(N, \P | \P_0, \varphi_0)$  and its integration interval. First of all, we note that the argument of the exponent of $P$ defined in Equation~\ref{eq:p_stokes} is the distance between the measured point with coordinate $(q_0,u_0)$ and the generic point on the plane $(q,u)$. Recalling Equation~\ref{eq:p} and that $\sigma = \sigma_q = \sigma_u =\sqrt{2/N}$, such a term can be rewritten as (see Figure~\ref{fig:exp_term}):
\begin{eqnarray}
  -\frac{(q-q_0)^2 + (u-u_0)^2}{2\sigma^2} &=& -\frac{ N\mu^2}{4}\left\{\P^2 + \P_0^2 - 2\P \P_0\cos\left[2(\varphi-\varphi_0)\right]\right\} = \notag\\
  & = & -\frac{ N\mu^2}{4} \delta_\P^2 .
  \label{eq:exp_term}
\end{eqnarray}
Here $\delta_\P = \P^2 + \P_0^2 - 2\cos\left[2(\phi-\phi_0)\right]$ is essentially the difference in degree of polarization between the measured point $(q_0,u_0)$ and the generic point $(q,u)$.

The integration interval is trivial to define in the $(q,u)$ plane, as it is a circle with center the measured point$(q_0, u_0)$ and radius $\mu\delta_\P$. Such a region is more easy to integrate if we pass from variables $(q,u)$ to $(\delta_\P, \psi)$ (see Figure~\ref{fig:exp_term}). Then:

\begin{equation}
	\begin{cases}
		q = q_0 + \mu\delta_\P\cos\psi \\
		u = u_0 + \mu\delta_\P\sin\psi \\
	\end{cases} 
	\;\Rightarrow \;
	\d q \d u = \mu^2\delta_\P \d \delta_\P \d\psi .
	\label{eq:jacobian}
\end{equation}

With the help of Equation~\ref{eq:exp_term} and Equation~\ref{eq:jacobian}, Equation~\ref{eq:confidence} can eventually be rewritten as:
\begin{equation}
	C = \frac{N}{4\pi}\int_0^{2\pi} \d\psi \int_0^{\delta_\text{max}}   \exp{\left(-\frac{N\mu^2}{4}\delta_\P^2\right)} \mu^2 \delta_P \d\delta_\P.
\end{equation}
The integration is now trivial as $\delta_\P$ does not depend on $\psi$ and the result is \cite{Weisskopf2010}:
\begin{equation}
	\delta_\text{max} = \sqrt{-\frac{4\ln(1-C)}{N\mu^2}} .
	\label{eq:delta_max}
\end{equation}

We have now completely characterized the confidence region in the $(q,u)$ plane as a circle with radius $\mu\delta_\text{max}$ and described with the parametric variable $\psi$ running in the interval $[-\pi, \pi]$. To transform such a region in $(\P,\varphi)$ plane we first note that (see Figure~\ref{fig:exp_term}):
\begin{equation}
	\begin{cases}
		q = \mu\P_0\cos(2\varphi_0) + \mu\delta_\text{max}\cos\psi \\
		u = \mu\P_0\sin(2\varphi_0) +\mu\delta_\text{max}\sin\psi\\
	\end{cases} 
\end{equation}

and then \cite{Weisskopf2010, Strohmayer2013}:
\begin{equation}
	\begin{cases}
		\P = \frac{\sqrt{q^2 + u^2}}{\mu} = \sqrt{\P_0^2 + \delta_\text{max}^2+2\P_0\delta_\text{max}\cos(\psi-2\varphi_0)} \\
		\varphi = \frac{1}{2}\arctan(\frac{u}{q}) = \frac{1}{2}\arctan\frac{\P_0\sin(2\varphi_0) + \delta_\text{max}\sin\psi}{\P_0\cos(2\varphi_0) +\delta_\text{max}\cos\psi}
	\end{cases} .
	\label{eq:parametric_function}
\end{equation}

\begin{figure}
	\centering
	\includegraphics[width=0.7\textwidth]{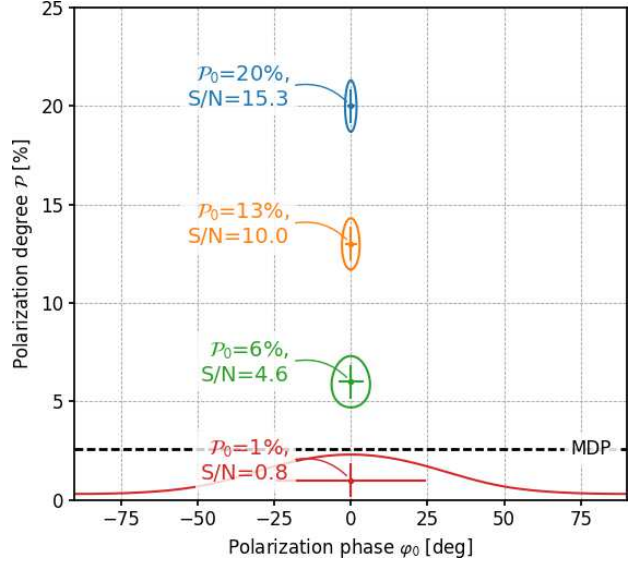}
	\caption{Example of visualization of a polarization measurement expressed in degree and angle, as a function of its signal-to-noise (S/N). The confidence level is 68.3\%, $N=300,000$, $\mu=0.3$ and $\varphi_0=0$. Error bars are derived with the ``1-D'' treatment.}
	\label{fig:pol_measurement}
\end{figure}

Equation~\ref{eq:parametric_function} is the system of parametric functions, with variable $\psi$, which describes the region on the $(\P,\varphi)$ plane at a confidence level $C$, which determines the value $\delta_\text{max}$ together with the number of collected counts $N$. Such a region is plotted in Figure~\ref{fig:pol_measurement} for a confidence level of 68.3\% for a decreasing significance of the measurement. While the region is essentially elliptical when the measurement has a high signal-to-noise (S/N), its contour tends to elongate when the S/N is low and all angles are possible when the measurement is not significant.

The maximum and minimum values of the polarization degree and angle inside the contour are the boundaries over which they vary at the selected confidence level. It is useful to calculate them when $C=68.3\%$, as this is the statistical uncertainty usually quoted with the measurement. For $\P$ the calculation is trivial, as it varies in the interval $[\P_0-\delta_\text{max}, \P_0+\delta_\text{max}]$ as $\psi$ varies and then:
\begin{equation}
	\sigma_\P = \delta_\text{max}|_{C=68.3\%} \approx \frac{1}{\mu}\sqrt{\frac{4.60}{N}} .
	\label{eq:sigma_p}
\end{equation}
This remains valid as long as the measurement is significant at the selected confidence level, that is, $\P_0>\delta_\text{max}$; otherwise, lower bound for $\P$ is zero. 

Deriving the same interval for $\varphi$ requires some math. Parametric function describing $\varphi$ has two critical points which can be written as:
\begin{equation}
	\varphi_{1,2} = \varphi_0 \pm \frac{1}{2}\arctan{\frac{\delta_\text{max}}{\sqrt{\P_0^2-\delta_\text{max}^2}}} .
\end{equation}

The largest value between $\varphi_1$ and $\varphi_2$ is the maximum polarization angle (it changes with $\psi$), whereas the minimum is the smaller of these two numbers. Therefore:
\begin{equation}
	\sigma_\varphi = \frac{1}{2} \arctan\frac{\delta_\text{max}|_{C=68.3\%}}{\sqrt{\P_0^2-(\delta_\text{max}|_{C=68.3\%})^2}} .
	\label{eq:sigma_varphi}
\end{equation}

In case the measurement has a high statistical significance, $\P_0\gg\delta_\text{max}$ and Equation~\ref{eq:sigma_varphi} can be approximated with:
\begin{equation}
	\sigma_\varphi\approx \frac{1}{2}\frac{\delta_\text{max}|_{C=68.3\%}}{\P_0} = \frac{1}{2}\frac{\sigma_\P}{\P_0} .
	\label{eq:sigma_varphi_approx}	
\end{equation}

There is another possible approach to derive the statistical uncertainty on a polarization measurement, which is to simply propagate statistical uncertainty on the measured Stokes parameters. Using Equations~\ref{eq:p} and \ref{eq:phi}, 1-$\sigma$ uncertainty propagated in this way reads \cite{Strohmayer2013}:
\begin{eqnarray}
\sigma_\P^{\text{1D}} &=& \frac{1}{\mu}\sqrt{\frac{q^2\sigma_q^2}{q^2+u^2}+\frac{u^2\sigma_u^2}{q^2+u^2}} = \frac{1}{\mu}\sigma = \frac{1}{\mu}\sqrt{\frac{2}{N}} \label{eq:sigma_p_1d}\\
\sigma_\varphi^{\text{1D}} &=& \frac{1}{2(q^2+u^2)}\sqrt{u^2\sigma_q^2 + q^2\sigma_u^2} = \frac{1}{2\mu\P}\sigma = \frac{1}{2\mu\P}\sqrt{\frac{2}{N}} = \frac{\sigma_\P^{\text{1D}}}{2\P}
\label{eq:sigma_varphi_1d}
\end{eqnarray}
where we neglected the uncertainties on $I$ or on $\mu$, as they are supposed to be much smaller that $\sigma_q$ or $\sigma_u$, and we used the superscript ``1-D'' to distinguish the values found with this approach from the ones in Equation~\ref{eq:sigma_p} and \ref{eq:sigma_varphi}. It is important to stress that this procedure does assume that $\P$ and $\varphi$ are independent variables, which are not. Therefore, the uncertainty calculated in this ways has a different meaning with the respect to the value calculated above, even if in both cases we assumed a confidence level of 68.3\%. Equation~\ref{eq:sigma_p} and \ref{eq:sigma_varphi} provide the \emph{combined} interval for both the polarization degree, that is, we can say that our measurement constrained the polarization degree in the interval $\P_0\pm\sigma_\P$ and the polarization angle in $\varphi_0\pm\sigma_\varphi$. On the contrary, 1-D uncertainties provide such an interval for only one of them, leaving the other one unconstrained \cite{Strohmayer2013}. Then, one can say that polarization has a 1-$\sigma$ uncertainty equal to $\sigma_\P^{\text{1D}}$, but as the polarization varies in such an interval the polarization angle is left unconstrained. 

1-D uncertainties are shown together with ``combined'' ones in Figure~\ref{fig:pol_measurement}. If we compare Equations~\ref{eq:sigma_p} and \ref{eq:sigma_p_1d}, 1-D uncertainty on polarization degree is nearly ~40\% smaller than the corresponding ``combined'' one. This is true also for the polarization angle, as in both approaches the uncertainty on $\varphi$ is approximately proportional to $\sigma_\P$. While the decision to use combined or 1-D uncertainties should taken on a case-by-case basis, when one wants to use the full potential of polarization measurement, that is, use both the constrain on polarization degree and angle, the appropriate statistical approach is the combined one.

\section{Conclusions}

In this chapter we have presented how a polarization measurement obtained with a photoelectric polarimeter is processed to derive the polarization degree and angle. Starting from the analysis of the raw data provided by this kind of instrument, that is, the image of the track of the photoelectron in a medium, we discussed different methods to derive the measured value and its statistical uncertainty, and to evaluate the credibility of the measurement. 

The intent was to collect in a single reference the basic information that a user needs to know to use data acquired with photoelectric polarimeters. This is timely, as the interest on this topic is rapidly increasing: at the time of writing, the Imaging X-ray Polarimetry Explorer (IXPE) has been just launched and it is expected to detect polarization from tens of sources of different classes. Before IXPE, PolarLight, which is a cubesat based on the same photoelectric polarimeter on-board IXPE, has already provided fresh data \cite{Feng2020}. In the next future, the enhanced X-ray Timing and Polarimetry mission \cite{Zhang2019} will expand IXPE results with a larger collecting area and simultaneous spectral and timing observation. 

All these opportunities have renewed the interest in the development of data analysis tools, which, however, are still discussed in specific research papers. Here we tried to collect and present the results with a uniform notation and approach, typically preferring  a simplified (and easier to follow) approach to a more rigorous treatment. The interested reader can find the latter in the original papers cited in the text. 

Finally, it is worth mentioning that the discussion presented here is largely shaped on the analysis tools developed for the IXPE mission. In fact, IXPE has been the first mission for which a complete set of analysis tools had to be distributed for the general user, and this has pushed the team to explore and tackle all the issue along the path, from data reduction to statistical treatment of the measurement. At the same time, IXPE example also guided the approximations done throughout this chapter, which are typically adequate for the analysis at its level of sensitivity.

\section{Acknowledgments}

The autor acknowledges the IXPE team for providing much of the material presented in this chapter.

\bibliography{References.bib}   
\bibliographystyle{elsarticle-num}

\end{document}